\documentclass[letterpaper,10pt]{article} 

\usepackage{opticameet3} 
\usepackage{xcolor}
\usepackage{soul}
\newcommand\authormark[1]{\textsuperscript{#1}}

\usepackage{amsmath,amssymb}
\usepackage[colorlinks=true,bookmarks=false,citecolor=black,urlcolor=black, linkcolor=black]{hyperref} 

\begin{document}

\title{Dual-Wavelength $\phi$-OFDR Using a Hybrid-Integrated Laser Stabilized to an Integrated SiN Coil Resonator}


\author{
    Mohamad Hossein Idjadi\authormark{1,*}, Stefano Grillanda\authormark{1},
    Nicolas Fontaine\authormark{1}, Mikael Mazur\authormark{1}, Kwangwoong Kim\authormark{1}, Tzu-Yung Huang\authormark{1}, Cristian Bolle\authormark{1}, Rose Kopf\authormark{1}, Mark Cappuzzo\authormark{1}, Kaikai Liu\authormark{2}, David A. S. Heim\authormark{2}, Andrew Hunter\authormark{2}, Karl D. Nelson\authormark{3}, and Daniel J. Blumenthal\authormark{2}
}

\address{\authormark{1} Nokia Bell Labs, Murray Hill, NJ 07974, USA, 
        \authormark{2} Department of Computer and Electrical Engineering, University of California at Santa Barbara, Santa Barbara, CA 93106, USA, 
        \authormark{3} Honeywell Aerospace, Plymouth, MN 55441, USA}
\email{\authormark{*}e-mail: mohamad.idjadi@nokia-bell-labs.com} 

\begin{abstract}
We demonstrate dual-wavelength distributed acoustic sensing over 37 km of standard single-mode fiber using $\phi$-OFDR, utilizing a scalable hybrid-integrated dual-wavelength laser chip frequency-locked to a high-Q integrated SiN coil resonator.
\end{abstract}
\section{Introduction}
Lasers with low phase noise, long-term stability, and high spectral purity are critical components in distributed acoustic sensing (DAS) of fiber optic cables \cite{ding2018distributed}. Among various fiber sensing architectures, phase-sensitive optical frequency domain reflectometry ($\phi$-OFDR) has garnered significant attention due to its high spatial resolution and moderate processing bandwidth \cite{yao2020integrated}. In particular, dual-wavelength $\phi$-OFDR systems, where two optical frequencies are locked to a common reference, have been developed to overcome limitations of earlier systems, improving both the dynamic range and signal-to-noise ratio (SNR) in phase or strain measurements of a fiber cable\cite{yang2023interval}. While there has been notable progress in the development of ultra-narrow linewidth lasers \cite{alkhazraji2023linewidth} and advancements in high-Q optical resonators and cavities \cite{liu2023integrated}, achieving an integrated optical frequency system that is CMOS-compatible, supports multi-wavelength operation, delivers milliwatt-level power per comb line, and maintains an ultra-narrow linewidth remains a significant challenge.

Here, we experimentally demonstrate the use of a scalable hybrid-integrated dual-wavelength laser, frequency-locked to a common high-Q integrated silicon nitride (SiN) coil resonator, for dual-wavelength DAS over 37 km of standard single mode fiber (SMF) using the $\phi$-OFDR technique. Our scalable approach provides a promising pathway toward a fully integrated, stable laser system for DAS interrogators.

\begin{figure}[b]
\centering
\includegraphics[width=\linewidth]{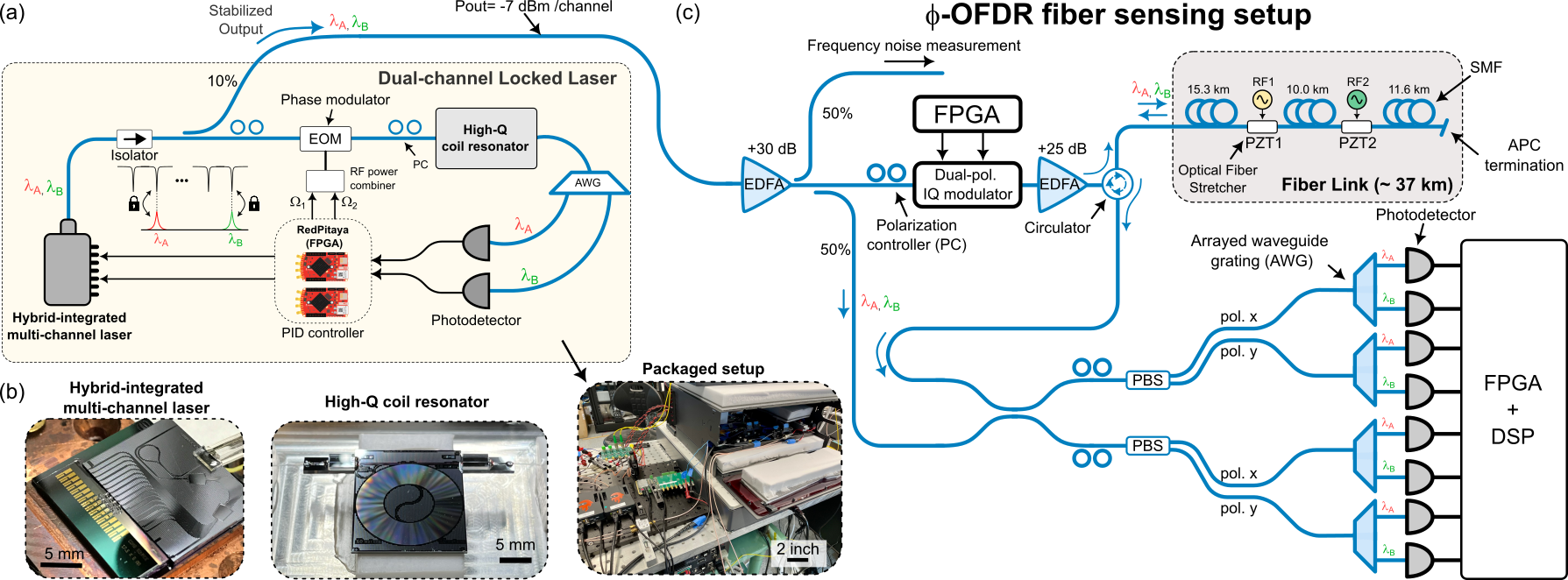}
\caption{\small \textbf{(a)} Dual-channel frequency-locked laser system. \textbf{(b)} Photograph of the hybrid-integrated laser and of the high-Q integrated SiN coil resonator, and the packaged setup. \textbf{(c)} Setup for $\phi$-OFDR DAS measurements. }
\label{fig1_diag}
\end{figure}
\section{Dual-channel frequency stabilized hybrid integrated laser}
Figure \ref{fig1_diag}(a) shows the block diagram of the dual-wavelength frequency stabilized hybrid integrated laser. The Pound-Drever-Hall (PDH) locking technique \cite{drever1983laser} is used to simultaneously lock the frequency of two channels ($\lambda_A, \lambda_B$) of the hybrid-integrated laser to two resonances of the integrated SiN coil resonator. For frequency noise analysis, a fully stabilized fiber frequency comb (see ref. \cite{idjadi2024hybrid}) is mixed with the hybrid-integrated laser output, and the RF beat notes ($f_{b1}, f_{b2}$) are digitized and processed. Details of the frequency locking and beat note measurement are explained in ref. \cite{idjadi2024hybrid}. Figure \ref{fig1_diag}(b) shows photographs of the hybrid-integrated multi-channel laser and the high-Q SiN coil resonator, both of which are packaged for enhanced acoustic shielding. 

The hybrid-integrated multi-channel laser consists of four III-V-based reflective semiconductor optical amplifiers and a passive silica-on-silicon chip, which integrates an arrayed waveguide grating (AWG) and a Sagnac loop. The two channels operate at 1533.51 nm and 1535.83 nm with intrinsic linewidth of $\sim$1 kHz, both frequency-locked to the SiN coil resonator, with each channel capable of delivering a maximum optical power of $\sim$7 mW. The laser chip has a footprint of 17.5 $\times$ 20 mm$^2$. The 10-meter CMOS-compatible SiN coil resonator, with a Q-factor of 64.4 million at 1535 nm, has a footprint of 26 $\times$ 21 mm$^2$. Additional details can be found in refs. \cite{idjadi2024hybrid, liu2023integrated, grillanda2022hybrid}.

\section{Experimental results}
\begin{figure}[t]
\centering
\includegraphics[width=\linewidth]{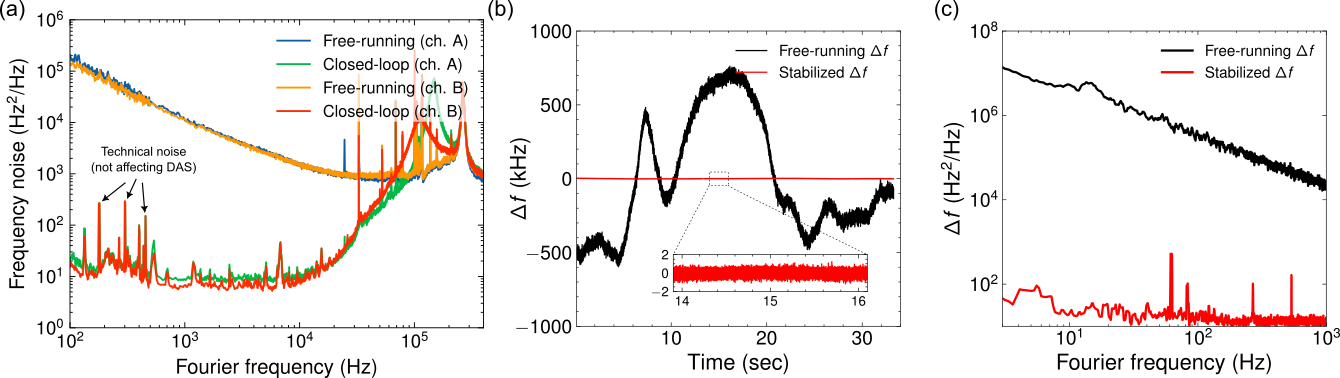}
\caption{\small \textbf{(a)} The frequency noise PSD of free-running and locked channels A and B. Technical noise spikes are related to the fiber frequency comb reference and do not affect the DAS measurement. \textbf{(b)} The difference in beat note frequencies between the two laser channels ($\Delta f$) in time-domain. \textbf{(c)} The PSD of $\Delta f$.}
\label{fig2_lock}
\end{figure}

\begin{figure}[b]
\centering
\includegraphics[width=\linewidth]{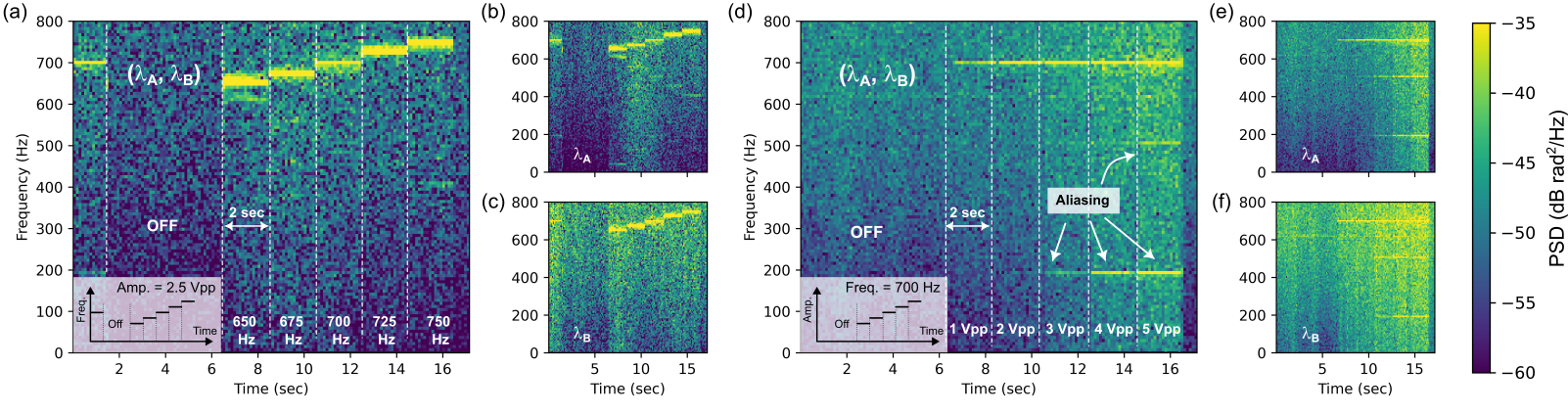}
\caption{\footnotesize \textbf{(a)} Spectrogram of the PSD of the phase difference $\Delta \phi_{AB}$ between $\lambda_A$ and $\lambda_B$ at PZT1 location for variable-frequency excitation, with sinusoid frequency changing from 650 Hz to 750 Hz every 2 seconds. \textbf{(b, c)} Spectrogram of the PSD of phase for individual laser channels in (a). \textbf{(d)} Spectrogram of the PSD of $\Delta \phi_{AB}$ for variable-amplitude excitation, with signal amplitude varying from 1 V$_{pp}$ to 5 V$_{pp}$ every 2 seconds. \textbf{(e, f)} Spectrogram of the PSD of phase for individual laser channels in (d). Figures (a-f) share the same x-axis, y-axis, and colorbar.}
\label{fig3_ofdr}
\end{figure}
Figure \ref{fig1_diag}(c) illustrates the $\phi$-OFDR fiber sensing setup. The stabilized laser channels are first amplified using an Erbium-Doped Fiber Amplifier (EDFA); then, half of the amplified light is directed toward the laser frequency noise and beat note measurement setup, the other half is further split, with one portion being modulated by a dual-polarization I/Q modulator driven by an FPGA, while the other portion serves as the local oscillator (LO) in the self-heterodyne receiver setup. The FPGA is set to generate linearly chirped pulses with 250 MHz bandwidth, corresponding to a minimum theoretical spatial resolution of about 40 cm \cite{mazur2023advanced}.
The output from the dual-polarization I/Q modulator is amplified and then transmitted through the fiber link via an optical circulator. The fiber link consists of approximately 37 km of SMF, with two fiber stretchers, based on piezoelectric transducer tubes (PZTs), placed 10 km apart along the length of the link. The Rayleigh back-scattered laser signals are mixed with the LO signal using a self-heterodyne receiver. Polarization beam splitters (PBS) and fiber coupled AWGs are employed to separate the two polarizations and wavelength channels. An array of eight single-ended photodetectors (PDs) capture the optical signals, while the FPGA digitizes and a GPU processes the data.

Figure \ref{fig2_lock}(a) shows the power spectral density (PSD) of the frequency noise for the two laser channels in both free-running and stabilized conditions, with 40 dB of frequency noise suppression achieved for both channels. Figure \ref{fig2_lock}(b) illustrates the beat note frequency difference between the two laser channels (i.e., $\Delta f = f_{b1} - f_{b2}$) recorded over 34 seconds. When the laser channels are locked, the root mean square (rms) of random frequency fluctuations is reduced by more than 25 dB. Figure \ref{fig2_lock}(c) shows the PSD of the frequency difference between the laser channels ($\Delta f$), demonstrating more than an order of magnitude improvement compared to ref. \cite{idjadi2024hybrid}.

Figure \ref{fig3_ofdr} presents the $\phi$-OFDR experimental results with both laser channels frequency locked. Figure \ref{fig3_ofdr}(a) depicts the spectrogram of the PSD of the phase at the location of PZT1, where the phase difference between the two locked laser channels (i.e. $\Delta \phi_{AB}$) is measured. As shown in the inset in Fig. \ref{fig3_ofdr}(a), a sinusoidal waveform with a fixed amplitude of 2.5 V$_{pp}$ and a variable frequency, ranging from 650 Hz to 750 Hz, is applied to PZT1. Figures \ref{fig3_ofdr}(b) and \ref{fig3_ofdr}(c) show the individual PSD of the phase for individual laser channels. Figure \ref{fig3_ofdr}(d) illustrates the measured spectrogram of the PSD of $\Delta \phi_{AB}$ when a sinusoidal waveform with a fixed frequency and a variable amplitude (ranging from 1 V$_{pp}$ to 5 V$_{pp}$) is applied to PZT1. Figures \ref{fig3_ofdr}(e) and \ref{fig3_ofdr}(f) display spectrogram of the PSD of the phase for each laser channel individually. From Figs. \ref{fig3_ofdr}(a-f), it is clear that using two locked laser channels to probe the fiber reduces common mode noise and enhances both the SNR and measurement sensitivity. Moreover, as shown in Fig. \ref{fig3_ofdr}(f), the background phase noise increases significantly with the rise in PZT modulation amplitude. It is worth noting that the sampling rate is 1.9 kHz and increasing the modulation amplitude leads to harmonic aliasing, causing the phase noise to fold back into the baseband, raising the noise floor for both laser channels, similar to what occurs in sampled systems \cite{arias2011noise}. However, since the two channels are correlated and frequency-locked to a common reference, subtracting the fiber phase noise suppresses the noise-folding effect.

\begin{figure}[t]
\centering
\includegraphics[width=\linewidth]{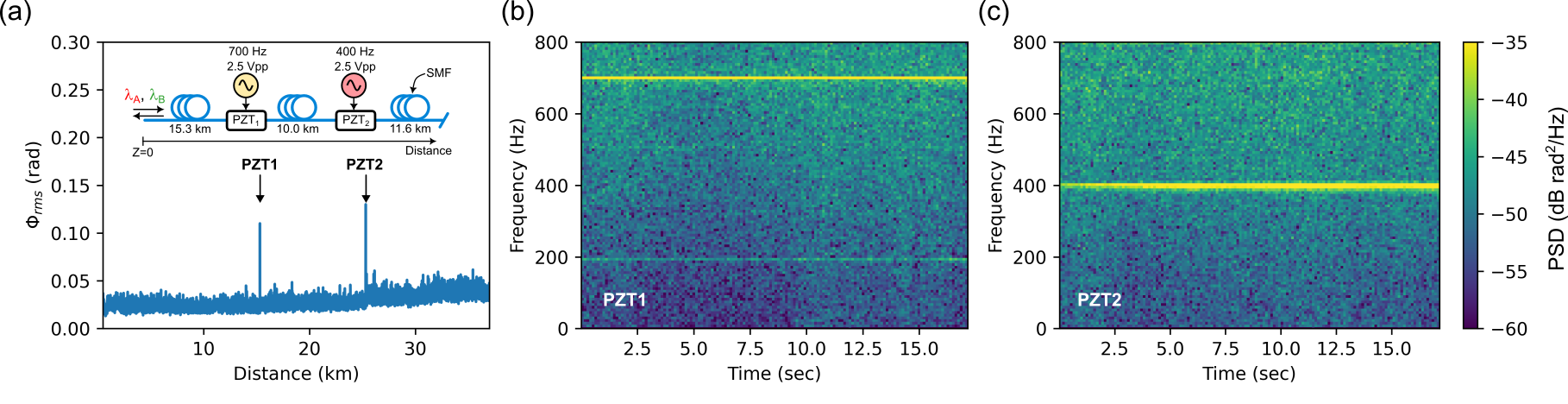}
\caption{\footnotesize  \textbf{(a)} Simultaneous electrical excitation at 700 Hz and 400 Hz applied to PZT1 and PZT2, respectively and measured rms DAS phase over a 37 km SMF link. \textbf{(b, c)} Spectrograms of $\Delta \phi_{AB}$ at PZT1 and PZT2, respectively.}
\label{fig4_ofdr}
\end{figure}

Figure \ref{fig4_ofdr}(a) presents the setup and the measured rms phase along the fiber link, revealing two distinct peaks at the locations of PZT1 and PZT2, corresponding to simultaneous excitations at 700 Hz and 400 Hz, respectively. Figures \ref{fig4_ofdr}(b) and \ref{fig4_ofdr}(c) display the spectrograms of the PSD of $\Delta \phi_{AB}$ at PZT1 and PZT2, respectively.

\section{Conclusion}
A frequency-stabilized dual-wavelength hybrid-integrated laser is developed to demonstrate DAS over a 37 km SMF link using dual-wavelength $\phi$-OFDR. The demonstrated interrogator can detect dynamic events with enhanced measurement sensitivity. This integration of scalable multi-channel laser technology with an ultra-low-loss SiN platform represents a significant advancement toward fully integrated, scalable, and stable laser systems for DAS interrogators.


\end{document}